\documentclass[aps,pra,twocolumn,showpacs,superscriptaddress,longbibliography]{revtex4-2}

\usepackage{graphicx}
\usepackage{dcolumn}
\usepackage{bm,amssymb}
\usepackage{braket}
\usepackage[normalem]{ulem}
\usepackage{xcolor} 
\usepackage[most]{tcolorbox}
\usepackage{booktabs}
\usepackage{diagbox}
\usepackage{soul}
\graphicspath{{figures/}}
\begin{document}

\title{Robust Rydberg facilitation via rapid adiabatic passage}
\author{Xinghan Wang}%
\affiliation{ 
Department of Physics and Astronomy, Purdue University, West Lafayette, IN 47907, USA
}%

\author{Yupeng Wang}%
\affiliation{ 
Department of Physics and Astronomy, Purdue University, West Lafayette, IN 47907, USA
}%

\author{Qi-Yu Liang}
\affiliation{ 
Department of Physics and Astronomy, Purdue University, West Lafayette, IN 47907, USA
}%
\affiliation{Purdue Quantum Science and Engineering Institute, Purdue University, West Lafayette, IN 47907, USA}

\date{\today}

\begin{abstract}
We propose and analyze a robust implementation of Rydberg antiblockade based on rapid adiabatic passage. Although Rydberg antiblockade offers key opportunities in quantum information processing and sensing, its sensitivity to position disorder and parameter imperfections has posed a central roadblock. By adiabatically sweeping across the interaction-shifted resonance, our approach is unaffected by realistic levels of disorder and parameter variations. As a straightforward application case, we show that it naturally gives rise to avalanche excitation growth in both one- and two-dimensional arrays. This avalanche process yields high gain with exceptionally low background, making it promising for rare-event detection. These results establish a practical route to robust Rydberg antiblockade dynamics, paving the way for future experimental and technological applications.

\end{abstract}

\maketitle
\section*{Introduction}
Rydberg blockade~\cite{henriet2020quantum,graham2022multi,muniz2025high} arises when the interaction shift between two nearby Rydberg atoms is much larger than the excitation Rabi frequency. In this regime, the laser drive is resonant for creating the first excitation but detuned for the second, effectively suppressing multiple excitations. The result is a robust mechanism for generating entanglement and multi-qubit gates, since only the existence of a large interaction matters, not its exact value. By contrast, Rydberg antiblockade~\cite{su2020rydberg,chew2022ultrafast} refers broadly to regimes where simultaneous excitation of multiple atoms occurs in spite of interactions, either because the interaction is weak, comparable to the drive, or actively compensated by laser detuning.
Antiblockade has been proposed as a route to faster quantum logic gates free from blockade error and to implement SWAP operations without the overhead of decomposing them into multiple elementary gates~\cite{li2024high,shi2017rydberg,wu2021one}. Moreover, antiblockade enables entanglement generation~\cite{zhao2017robust}, provides avalanche amplification for sensing weak signals~\cite{nill2024avalanche} and constitutes a pivotal ingredient for kinetically-constrained lattice models that test the limits of our understanding of quantum thermalization and quantum matter~\cite{zhao2025observation,zhang2024quantum,liu2022localization,causer2020dynamics,zadnik2023slow}.

Despite the promising landscape spanning quantum primitives to fundamental physics, the stringent requirement of precise interaction control has hindered experimental progress in exploiting Rydberg antiblockade and accessing its associated opportunities. A primary challenge is positional disorder: finite atomic temperature translates into disorder in the interactions due to their strong distance dependence. Consequently, high-fidelity antiblockade protocols typically require advanced cooling techniques that are beyond the scope of typical tweezer-array experiments. Even with such techniques, the achieved fidelity often remains non-competitive. For example, while a recently proposed high-tolerance antiblockade SWAP gate predicts a fidelity of 0.955, a compiled SWAP gate, within the same platform, built from experimentally demonstrated~\cite{evered2023high,radnaev2025universal} CZ gates (0.995 fidelity) and single-qubit rotations (0.9997 fidelity) is expected to reach a fidelity of 0.982, significantly outperforming the former. 

In this work, we propose to circumvent this sensitivity by adiabatically sweeping through the interaction-shifted resonance, thereby enabling a robust implementation of the antiblockade mechanism. Specifically, we employ rapid adiabatic passage (RAP) to drive state transitions. RAP is a technique widely used for robust state preparation, which has also been proposed to generate entanglement and realize two-qubit gates~\cite{zhao2017robust,mitra2023neutral,xue2024high}. We compare with a resonant Rabi protocol to showcase its robustness to variations in driving parameters and, more importantly, position disorders. Although adiabaticity is typically associated with longer evolution time, our study shows that under realistic constraints this is not necessarily the case. As an example application, we show that this mechanism can be leveraged to trigger avalanche Rydberg excitations, enabling amplified readout of a single initial Rydberg excitation. If such an initial excitation is generated as part of a sensing process, for example, as a consequence of detecting a microwave or terahertz photon~\cite{nill2024avalanche}, this low-dark-count, high-gain amplifier, may assist a variety of rare-event detection, including dark-matter searches~\cite{graham2024rydberg,engelhardt2024detecting}.



\section*{Results}

\subsection*{Model} 
Consider a 1D array of $N$ sites. Each site is occupied by a single atom. A driving laser field with time-dependent detuning $\Delta(t)$ and site-specific, time-dependent Rabi frequency $\Omega_j(t)$ couples the ground state $\ket{0}_j$ to the Rydberg state $\ket{1}_j$ at site $j$. The single-body contribution to the Hamiltonian is given by
\begin{align}
H_{\text{site}}(t) &= \frac{1}{2} \sum_{j=1}^{N} 
 \Omega_j(t) \sigma_j^{x} 
 - \frac{\Delta(t)}{2} \sum_{j=1}^{N}\sigma_j^{z} 
\end{align}
where $\sigma_{j}^{x} = \lvert 1 \rangle_{j j} \langle 0 \rvert 
+ \lvert 0 \rangle_{j j} \langle 1 \rvert$ and $\sigma_{j}^{z} = \lvert 1 \rangle_{j j} \langle 1 \rvert 
- \lvert 0 \rangle_{j j} \langle 0 \rvert$.

The van der Waals interaction between Rydberg atoms scales as $V_r=C_6/|\bm{r}|^6$, where $C_6$ is the van der Waals interaction coefficient (along the array direction if anisotropic). The resulting nearest-neighbor interaction in a chain with spacing $r$ is
\begin{align}
H_{\text{NN}} &= V_{r} \sum_{j=1}^{N-1} n_jn_{j+1} 
\end{align}
where $n_i=\ket{1}_{ii}\bra{1}$. We operate in the ``facilitated" or ``antiblockade" regime:  The ground-to-Rydberg transition is driven off-resonantly. The kinetic constraint ensures that resonance is achieved if and only if one of the nearest neighbors is in the Rydberg state. To execute RAP, we slowly ramp the laser detuning $\Delta(t)$ through this interaction-shifted resonant condition, $\Delta_0=V_r$.

\begin{figure}[ht]
\centering
\includegraphics[width=0.47\textwidth]{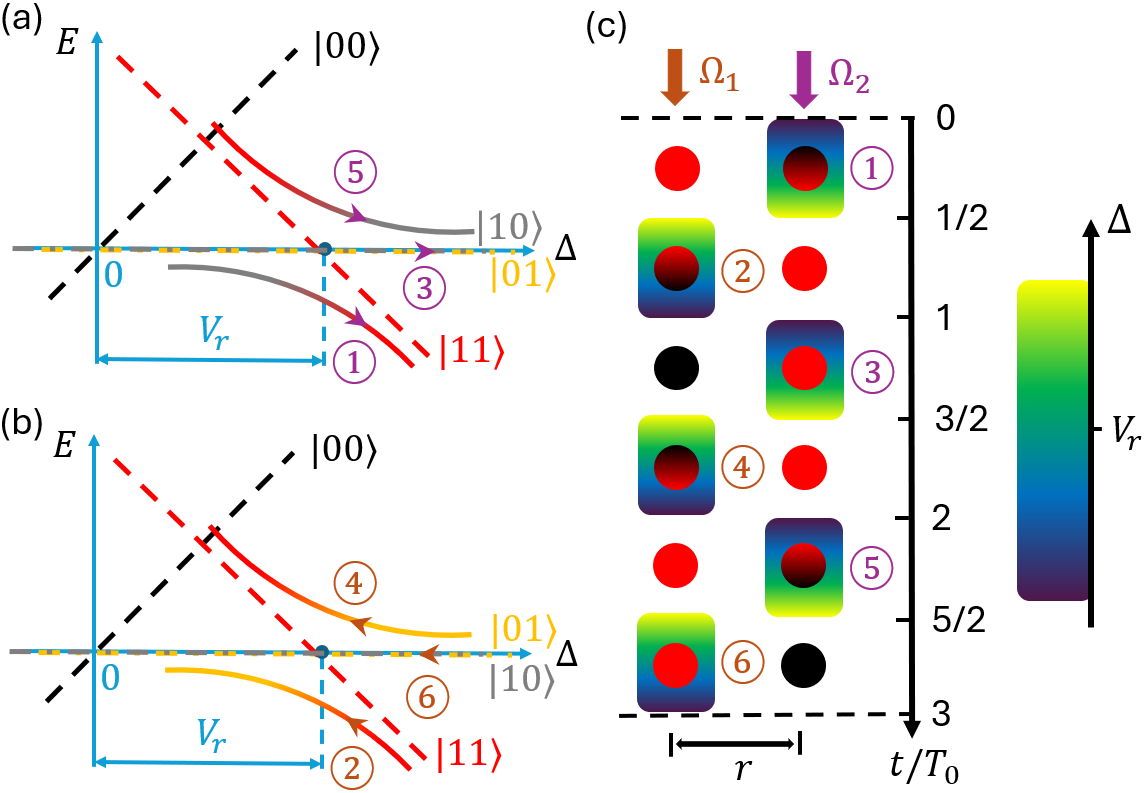}
\caption{Robust Rydberg facilitation in a two-site array. (a) Sketch of energy diagrams as a function laser detuning $\Delta$ for driving atom 2. (b) Same for driving atom 1. The dashed lines indicate bare states (red $\ket{11}$, grey $\ket{10}$, orange $\ket{01}$ and black $\ket{00}$). (c) Evolution of the Rydberg excitation under the periodic driving. \textcircled{1} $\ket{10}\rightarrow\ket{11}$, \textcircled{2} $\ket{11}\rightarrow\ket{01}$, \textcircled{3} stays in $\ket{01}$, \textcircled{4}  $\ket{01}\rightarrow\ket{11}$, \textcircled{5} $\ket{11}\rightarrow\ket{10}$ and \textcircled{6} stays in $\ket{10}$.}
\label{fig:energy level}
\end{figure}

\subsection*{Pulse sequence: two-atom demonstration}
First, we study a two-site case where the total Hamiltonian is $H=H_{\text{site}}(t)+H_{\text{NN}}$. Initially, a Rydberg atom is located in site 1 whereas site 2 is occupied by a ground-state atom. We alternatively drive the two atoms. When driving atom 2 with Rabi frequency $\Omega_2$, the initial state $\ket{10}$ is coupled to $\ket{11}$. The ramping step \textcircled{1} adiabatically follows the resulting upper dressed state (Fig.~\ref{fig:energy level}(a)). The second ramping step drives atom 1, coupling $\ket{11}\leftrightarrow\ket{01}$. By adiabatically sweeping the detuning $\Delta$ from above $V_r$ to below, the state is transferred to $\ket{01}$. In Fig.~\ref{fig:energy level}(c), the colored balls encode on-site populations: red denotes the Rydberg state $\ket{1}$, and black denotes the ground state $\ket{0}$. A red $\leftrightarrow$ black change marks an RAP. The rectangles indicate which atom is driven, with the color gradient showing the sweep direction of $\Delta$ from top to bottom as time evolves. Steps \textcircled{3} and \textcircled{6} do not result in any population transfer due to the energy penalty of interaction energy $V_r$. These steps would lead to population transfer if the detuning $\Delta$ is swept across 0 where the state $\ket{10}$ or $\ket{01}$ crosses state $\ket{00}$ in the energy diagram. The driving parameters complete one cycle $T_0$ in two steps, whereas the excitation hops between sites 1 and 2 with a period of $3T_0$.

We deploy the Allen-Eberly adiabatic passage scheme~\cite{osti_7365050,da2022shortcuts}. The Rabi frequencies and detuning evolve as follows within one cycle $0\le t<T_0$. For the first half-cycle ($0\le t<T_0/2$):
    \begin{subequations}
    \begin{align}
\Omega_{1}(t) &=0, \quad
\Omega_{2}(t) = \Omega_{0}\,\text{sech}\!\left[\frac{\pi\!\left(t-T_0/{4}\right)}{T_0/8}\right], 
 \\
\Delta(t) &= 
\Delta_0+\left(\frac{\beta^{2}T_{0}}{8\pi}\right)\tanh\!\left[\frac{\pi (t-T_0/4)}{T_0/8}\right]. 
\end{align}
\label{eq:1}
\end{subequations}
For the second half-cycle $T_0/2\le t<T_0$:
    \begin{subequations}
    \begin{align}
\Omega_{1}(t) &=\Omega_{0}\,\text{sech}\!\left[\frac{\pi\!\left(t-3T_0/{4}\right)}{T_0/8}\right],\quad
\Omega_{2}(t) = 0,\\
\Delta(t) &= 
     \Delta_0-\left(\frac{\beta^{2}T_{0}}{8\pi}\right)\tanh\!\left[\frac{\pi (t-3T_0/4)}{T_0/8}\right].  
\end{align}
\label{eq:2}
\end{subequations}
The sweep direction of the detuning can be swapped between the first and second half-cycle without affecting any population transfer. From an experimental perspective, this protocol requires site-resolved laser intensity ramping, but the detuning ramping is global, which is straightforward to implement.

\subsection*{Robustness to imperfections}
\begin{figure}[ht]
\centering
\includegraphics[width=0.43\textwidth]{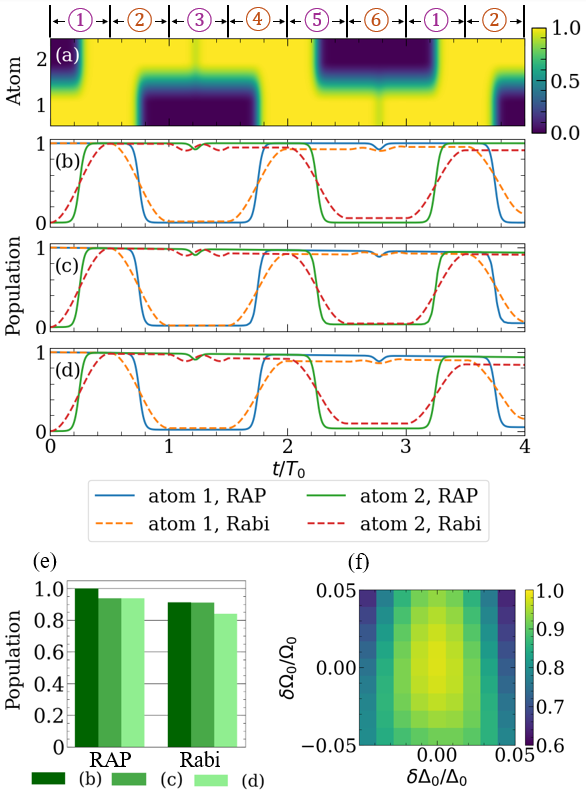}
\caption{Comparison of sensitivity to imperfections between RAP and Rabi schemes. (a) Rydberg population evolution under the pulse sequences of Fig.~\ref{fig:energy level} and Eqs.~(\ref{eq:1})(\ref{eq:2}), without position disorder or decay. Evolution subject (b) to position disorder, (c) to Rydberg-state decay $\Gamma$, and (d) to both effects. Parameters are $\Omega_0/(2\pi)=\{10.7,0.34\}$~MHz, $\sigma_r=\{76,34\}$~nm, $\overline{\delta V_{r}}/(2\pi)=\{1.37,0.02\}$~MHz, $V_r/(2\pi)=\{20,1.1\}$~MHz and $r=\{6.4,10.4\}~\mu$m for the RAP and Rabi schemes, respectively. (e) Final Rydberg population of atom 2 at $t=4T_0$, extracted from panels (b-d). (f) Rydberg population of atom 2 at $t=4T_0$ for the Rabi scheme as a function of detuning error $\delta\Delta_0$ and Rabi-frequency error $\delta\Omega_0$, without disorder or decay. }
\label{fig:disorder_decay}
\end{figure}

To evaluate the robustness of our scheme, we consider implementation in $^{87}$Rb tweezer arrays and compare RAP with a resonant Rabi scheme based on sequential $\pi$ pulses. We focus on one-photon excitation to a P Rydberg state. At room temperature (300 K), the decay of Rydberg states is dominated by blackbody-radiation–induced transitions. In the more common two-photon excitation scheme, admixture of a short-lived intermediate state can significantly increase the decay rate. Because RAP strongly suppresses the impact of disorder and sensitivity to driving parameters, Rydberg-state decay becomes the primary limitation. We therefore adopt the one-photon excitation to minimize decay and highlight the strength of RAP. The case of a two-photon excitation to an $S$ or $D$ Rydberg state, where RAP remains advantageous, is discussed in the Supplementary Information.

To simulate the dynamics, we use QuTiP~\cite{johansson2012qutip,lambert2024qutip}. 
The Rydberg decay is described by the Lindblad superoperator:
\begin{align}
\mathcal{L}_{decay}(\rho) &= \frac{\Gamma}{2}\sum_{j=1}^{N}  \left( 2 \sigma^-_j \rho \sigma^+_j - \sigma^+_j \sigma^-_j \rho - \rho \sigma^+_j \sigma^-_j  \right),
\end{align}
where $\sigma_i^+=\ket{1}_{ii}\bra{0}$ and $\sigma_i^-=\ket{0}_{ii}\bra{1}$.
The density matrix $\rho$ evolves according to the Lindblad master equation:
\begin{align}
    &\frac{d\rho}{dt}=i[\rho,H]   +\mathcal{L}_{decay}(\rho).
    \label{eq: master equation}
\end{align}


For the Rabi scheme, the laser detuning is set at the interaction-shifted resonance $\Delta_0$. Each step in Fig.~\ref{fig:energy level} is implemented by applying a $\pi$ pulse of duration $T_0/2=\pi/\Omega_0$ with constant Rabi frequency $\Omega_0$. The sequence of driven atoms and the intended state transfers are identical to those in the RAP scheme. 

A key challenge for the Rabi scheme is position disorder, which introduces static deviations in the interaction energy~\cite{wang2025directional,zhao2025observation,marcuzzi2017facilitation,valencia2024rydberg}, characterized by 
\begin{equation}
    \overline{\delta V_{r}}\approx6 |V_{r}|^{7/6}\frac{\sqrt{2}\sigma_r}{|C_6|^{1/6}},
    \label{eq: interaction disorder}
\end{equation}  
where $\sigma_r$ denotes the root-mean-square (rms) radius of the atomic position distribution along the array direction. 
Although going to higher principal quantum number $n$ and reducing interaction $V_r$ alleviate the impact of position disorder, the required interatomic spacing can become too large to fit a reasonable number of sites within the imaging field of view. Moreover, high $n$ states are more vulnerable to stray electric fields~\cite{panja2024electric,Ananddualspeciesrydberg,wilson2022trapping}. Taking 70P$_{3/2}$ as a reasonable choice, we conducted a limited parameter search to find workable parameters that balance disorder sensitivity with decay. This yields $V_r/(2\pi)=1.1~$MHz and $\Omega_0/(2\pi)=0.34~$MHz, accounting for Rydberg decay rate $\Gamma/(2\pi)=0.839~$kHz and position disorder corresponding to $\sigma_r=34~$nm. The disorder is modeled by averaging over 50 realizations, in which each atom’s displacement from its ideal position is randomly sampled from a Gaussian distribution with standard deviation $\sigma_r$.

This level of position distribution can be achieved by Raman sideband cooling~\cite{10.21468/SciPostPhys.10.3.052,jenkins2022ytterbium} or narrow-line cooling~\cite{blodgett2025narrow}, approaching the motional ground state~\cite{PhysRevA.110.053518}. To evaluate performance under these conditions, we simulate a two-atom array driven for $t=4T_0$. Panel (a) shows the evolution of the Rydberg excitation, initially located at site 1, following the RAP scheme, without decay or disorder. An ideal Rabi scheme reproduces the same state transfers at the end of each step. Panels (b–d) illustrate how position disorder, decay, and their combination affect the dynamics, while panel (e) presents the corresponding final Rydberg population at site 2. Under our parameter choice, both effects comparably reduce its value from the ideal value of unity. We emphasize that we have assumed extreme cooling and very large atomic spacing to mitigate position disorder; a naive Rabi implementation would suffer much more severe degradation from position disorder.

For RAP, the protocol speed is set by $\Omega_0$, which is technically limited by available laser power. Currently, 3~W laser at the transition ($5S_{1/2}\leftrightarrow70P_{3/2}$) wavelength 297~nm is available, and future experiments may further enhance the Rabi frequency using a build-up cavity. In any case, laser power is a less fundamental limitation than the spread of the ground-state wavefunction. We choose $\Omega_0/(2\pi)=10.7~$MHz, which requires 2.6~mW per site for a beam waist ($1/e^2$ radius) of 1.5~$\mu$m.
With $T_0=3~\mu$s and $\beta/(2\pi)=7.6~$MHz, determined by the Allen-Eberly pulse shape and the chosen $\Omega_0$, we find that the final Rydberg population at site 2 is reduced only by the blackbody-radiation-limited Rydberg lifetime (Fig.~\ref{fig:disorder_decay} (b-e)). Position disorder has no noticeable effect at a much larger atomic position spread. The value $\sigma_r=76~$nm used in our RAP simulation can be achieved by a radial trap frequency of 91~kHz at a  temperature of 20~$\mu$K, readily accessible by a simple and efficient cooling method: $\Lambda-$enhanced grey molasses~\cite{evered2023high,rosi2018lambda}.

In addition to position disorder and decay, we also consider imperfections in the driving parameters, namely deviations of the detuning $\Delta_0$ by $\delta\Delta_0$ and of the Rabi frequency $\Omega_0$ by $\delta\Omega_0$. As shown in Fig.~\ref{fig:disorder_decay} (f), for the Rabi scheme, a $\pm 5\%$ change in both can reduce the final Rydberg population from unity to about 0.6 ($\delta\Omega_0>0$, 0.8 for $\delta\Omega_0<0$). The apparent trend that larger $\Omega_0$ performs worse than its smaller counterpart is coincidental: if the sequence included one more or one fewer step of excitation transfer, the comparison could reverse and the larger $\Omega_0$ would appear more favorable.
For the RAP scheme, we do not show a plot because it is essentially insensitive to any realistic driving parameter error: scanning the same relative variations $\delta\Delta_0/\Delta_0$ and $\delta\Omega_0/\Omega_0$ changes the final population by less than $2\times10^{-4}$.

\subsection*{Avalanche excitations: 1D chain}

\begin{figure}[ht]
\centering
\includegraphics[width=0.30\textwidth]{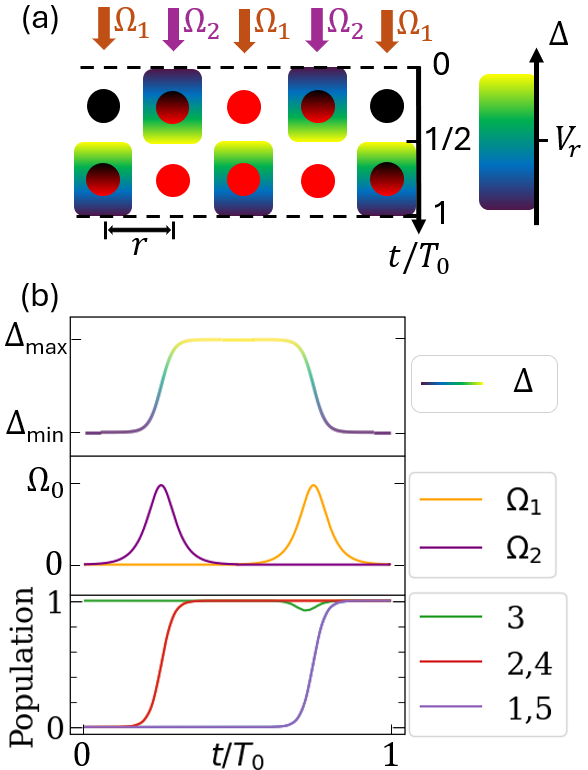}
\caption{Mechanism of avalanche excitation growth. After one driving period ($t=T_0$), all atoms are transferred to the Rydberg state. The detuning $\Delta$ is swept between $\Delta_{max,min}=\Delta_0\pm\beta^2T_0/(8\pi)$. The bottom panel of (b) shows the Rydberg population for each site: site 3 (green), sites 2 and 4 (red), and sites 1 and 5 (purple).}
\label{fig:topo_pump}
\end{figure}
Next, we extend the system from the two-atom array to an $N$-atom one-dimensional chain. We continue using the same parameters for the chain as for the two-atom case. We do not consider position disorder or imperfections of driving parameters as they do not affect the Rydberg population evolution. The driving of sites 1 and 2 naturally generalizes to all odd and even sites, respectively, while all other elements of the RAP protocol, including the Allen–Eberly pulse shapes, the cyclic sequence structure, the overall mechanism and parameter choices, remain unchanged. In this case, the site-dependent Rabi frequency takes the form:
\begin{equation}
  \Omega_j(t) = \begin{cases}\Omega_1(t)\quad\text{if $j$ mod $2=1$}\\
        \Omega_2(t)\quad\text{if $j$ mod $2=0$}
    \end{cases}
\end{equation}
Because of the sharp distance dependence of the interaction, the next-nearest-neighbor (NNN) coupling is $2^6=64$ times weaker than the nearest-neighbor (NN) term. Numerical checks confirm that including NNN interactions alters the results by less than one part in $10^4$, and can thus be neglected.

To illustrate how avalanche growth proceeds, we simulate a 5-atom chain and track how an initial seed excitation expands step by step (Fig.~\ref{fig:topo_pump}). Unlike in the two-atom case, here each half period leads to a state transfer. The excitation spreads symmetrically outward, adding one atom at each end in every step. Once an atom inside the cluster is excited, it is no longer driven down, because having two nearest-neighbor excitations shifts its resonance outside the RAP sweeping range. If the odd sites were driven in the very first step, no state transfer would occur in that step; from the second step onward, the growth is identical to the case where even sites are driven first. The cluster expands until it reaches the array boundary, after which it begins to shrink. To achieve a net gain of $2S+1$, $S$ steps are required, and the initial seed must have at least $S$ available sites on each side of the chain. 

\begin{figure}[ht]
\centering
\includegraphics[width=0.46\textwidth]{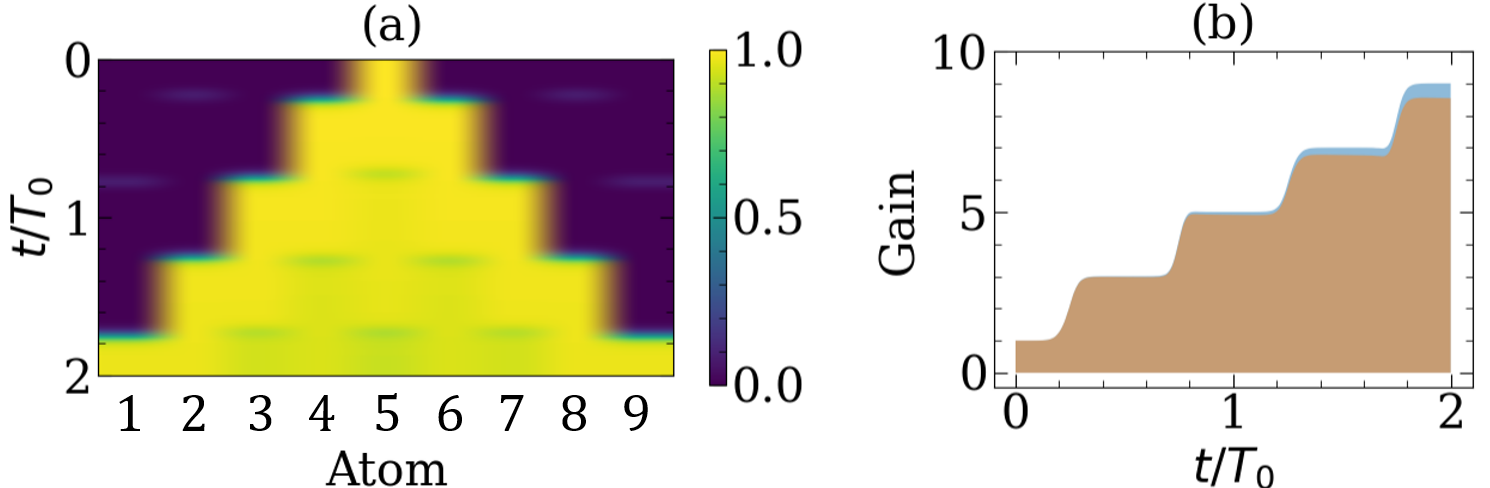}
\caption{Avalanche gain. (a) Rydberg population evolution. (b) Gain vs time: at $t=2T_0$, the no-decay value (blue) is short of the ideal value 9 by 0.003, while with decay (brow) it is 8.54. }
\label{fig:avalanche}
\end{figure}

Having established the avalanche mechanism, we now quantify its gain and noise performance.
Fig.~\ref{fig:avalanche} shows avalanche excitations starting from a single seed in the middle of a 9-atom chain. At the end of two driving period $t=2T_0$, the total number of excitations (gain) reaches 8.54. The reduction of gain from its ideal value $2S+1=9$ stems almost entirely from decay: without decay, the gain falls short of 9 only at the $10^{-3}$ level. By comparison, when starting with no excitation, the total number of excitations at this time is $7\times10^{-4}$. This background can be further suppressed by increasing $V_r$. For example, when $V_r=80$~MHz, the noise signal drops to $5\times10^{-6}$. This behavior is directly analogous to a single-photon detector, such as a photomultiplier tube or a single-photon avalanche diode: the system exhibits a high-gain, avalanche response triggered by a single seed excitation, while maintaining an exceptionally low dark count in its absence. Although the simulation uses a 9-atom chain, the result reflects the bulk behavior of longer arrays. In larger systems the avalanche can propagate across many more sites, leading to correspondingly greater gain, with the ultimate limit set by decay.

\subsection*{Gutzwiller mean-field approach}
However, we can only numerically solve the master equation (Eq.~\ref{eq: master equation}) for up to approximately 10 atoms. One common approach to this problem is to use Gutzwiller mean-field theory~\cite{jaksch1998cold}, which ignores the effects of correlations and assumes a product state: $\psi=\prod_j\psi_j$. Here, we combine the relative small effective Hilbert space of the Gutzwiller ansatz with quantum Monte Carlo wavefunction (QMCWF) simulation~\cite{casteels2018gutzwiller} to recover stochastic correlations generated by decay events, which are otherwise omitted if using a mean-field master equation.
Within this framework, the dynamics of each atom $j$ is governed by an effective single-site non-Hermitian Hamiltonian:
\begin{equation}
    H_{j}^{\mathrm{MF}}(t) = \frac{\Omega_{j}(t)}{2} \sigma_{j}^{x} - \frac{\tilde{\Delta}_{j}(t)}{2} \sigma_{j}^{z} -i\frac{\Gamma}{2}\sigma^+_j\sigma^-_j
\end{equation}
with effective detuning
\begin{equation}
\tilde{\Delta}_{j}(t) = \Delta(t) -  \sum_{k \in \mathcal{N}_j} V_{jk}\langle n_{k} \rangle(t)
\end{equation}
where $\langle n_k \rangle(t) = \bra{\psi_k(t)} n_k\ket{\psi_k(t)}$, $V_{jk} = \frac{C_6}{r_{jk}^6}$, $r_{jk}$ is the separation between sites $j$ and $k$, and $\mathcal{N}_j$ denotes neighbors of site $j$ that interact with it. The instantaneous rate for a quantum jump to occur on site $j$ is
\begin{equation}
   \frac{dp_j}{dt} =\Gamma\bra{\psi_j(t)} \sigma^+_j\sigma^-_j\ket{\psi_j(t)}
\end{equation}

\begin{figure}[ht]
\centering
\includegraphics[width=0.47\textwidth]{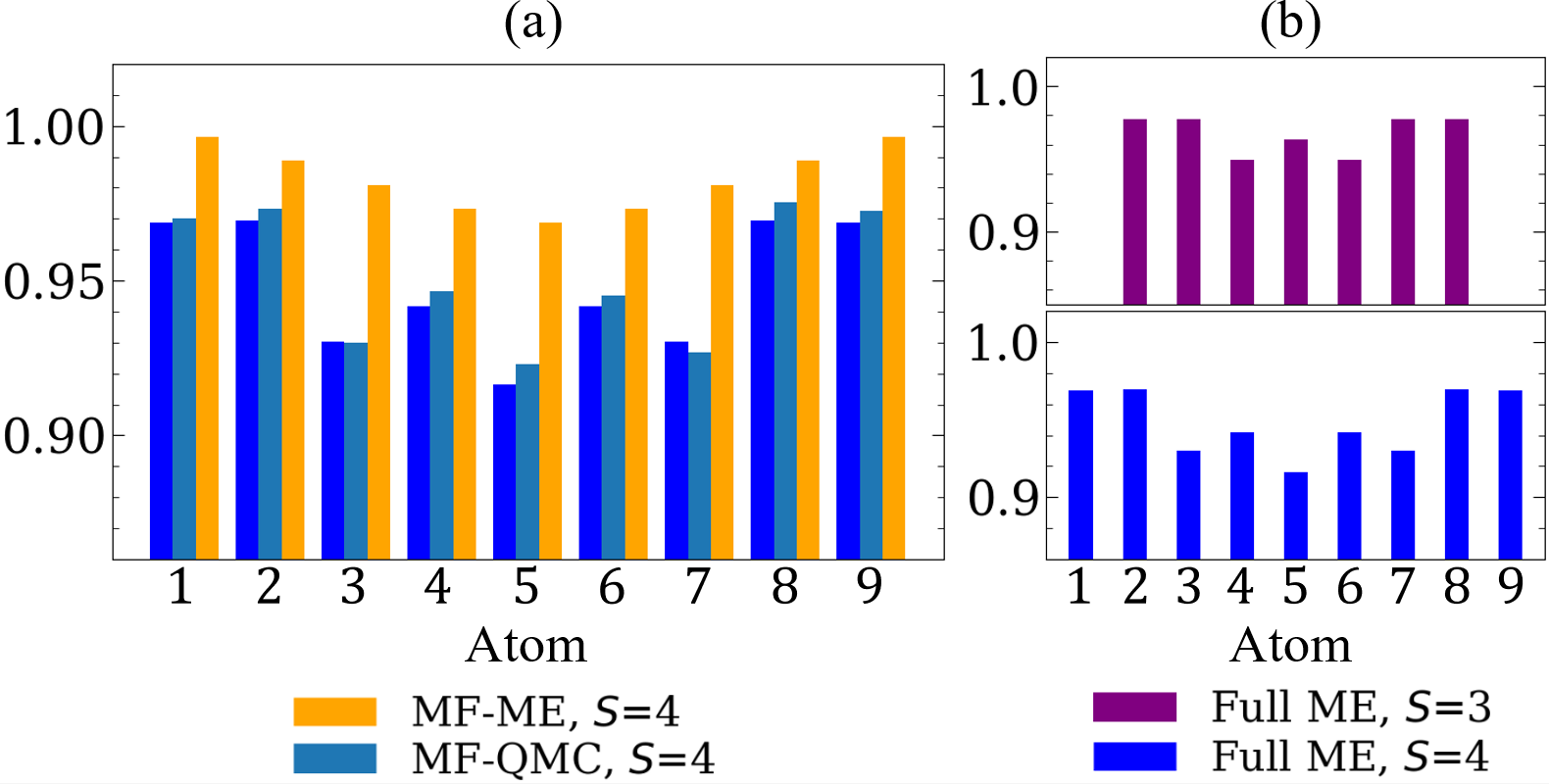}
\caption{Comparison of different simulation methods: full master equation (Full ME, blue), mean field with QMCWF (MF-QMC, sky blue) and mean-field master equation (MF-ME, orange). The blue bars in (a) and (b) show the same result as $t=2T_0$ in Fig.~\ref{fig:avalanche}. Panel (b) shows the Rydberg population pattern after $S=3$ steps (top) and after $S=4$ steps (bottom), illustrating two representative examples of the spatially oscillatory profile.}
\label{fig:method_comparison}
\end{figure}

We first validate this mean field with QMCWF approach by reproducing the 1D avalanche dynamics from Fig.~\ref{fig:avalanche}, finding agreement within 0.3\% using 700 quantum trajectories.
Fig.~\ref{fig:method_comparison}(a) compares the three approaches at $t=2T_0$, showing the site-resolved Rydberg populations under the same conditions as Fig.~\ref{fig:avalanche}: full master equation (Full ME), mean field with QMCWF (MF-QMC) and mean-field master equation (MF-ME). The mean-field master equation underestimates the impact of decay. It shows that earlier excitations are more likely to have decayed, leading to the lowest Rydberg population in the middle of the chain, but it does not capture the correlations. In this approach, decay only slightly reduce the average Rydberg population and hence the interaction with its neighbors, leaving the dynamics largely unchanged because it is robust to small imperfections. In reality, quantum correlations mean that (i) a decay event at the edge of a cluster can terminate excitation growth from that site, and (ii) a decay event in the bulk can seed the avalanche growth of a hole (ground-state atom) whenever only one excited neighbor is present. In other words, holes obey the same kinetic constraint as excitations, so the dynamics involves the correlated growth of both. For example, a decay at site 5 during step 2 would lead to hole growth to site 4 and 6 by the end of step 3.
 
\begin{figure}[ht]
\centering
\includegraphics[width=0.47\textwidth]{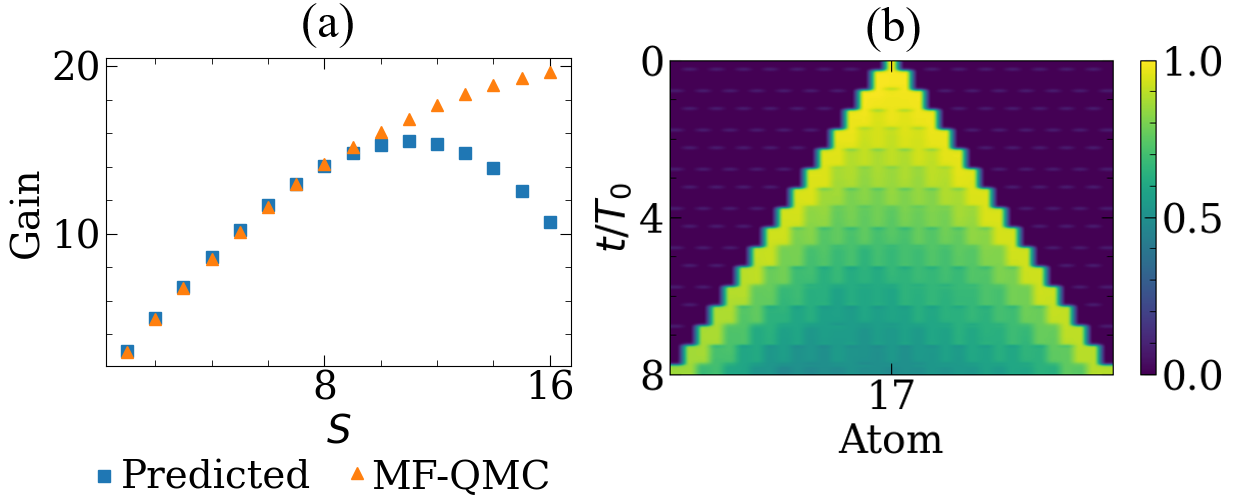}
\caption{Gain vs step number predicted by Eq.~\ref{eq: gain} (blue) and obtained from MF-QMC simulations (orange, averaged over 500 quantum trajectories). (b) Site-resolved population dynamics corresponding to the orange points in (a).}
\label{fig:exp_gain}
\end{figure}

Assuming small $\Gamma$ and restricting to at most one Rydberg decay event, we can explicitly include all possible hole–growth trajectories. This treatment reproduces the spatially oscillatory profiles seen in Fig.~\ref{fig:avalanche}(b): at a fixed step, the site populations within the excitation cluster exhibit an up–down modulation across neighboring sites. From one step to the next, the phase of this oscillation shifts by one site, so that the weakest-population sites switch parity (even sites at odd $S$, odd sites at even $S$). The same treatment further yields the analytical prediction for the gain after $S$ steps: 
\begin{equation}
    G=2S+1-\frac{\Gamma T_0}{2}(S^2+\frac{2}{3}S^3-\frac{2}{3}S-1),
    \label{eq: gain}
\end{equation}
as derived in the Supplementary Information, which also traces the oscillatory profiles back to hole–growth trajectories. The final gain for Fig.~\ref{fig:avalanche} from this equation is $G=8.57$, in good agreement with the full-master equation result. This equation also predicts the maximum gain of $G=15.1$ (Fig.~\ref{fig:exp_gain}(a)), reached at a finite step before decay reduces the excitation, assuming the array is sufficiently large that boundary effects do not constrain the dynamics.
To test this prediction, we extend the array size to 33 atoms and use MF-QMC to obtain gain vs step number $S$ (Fig.~\ref{fig:exp_gain}). Initially, the gain follows Eq.~(\ref{eq: gain}) closely. For larger $S$, however, the analytical expression underestimates the gain, because it neglects multiple decay events. In particular, a second decay event at the boundary of a hole can terminate its avalanche growth (see the Supplementary Information for further discussion of dynamics associated with multiple decay events).

\subsection*{Avalanche excitations: 2D square lattice}

\begin{figure}[ht]
\centering
\includegraphics[width=0.46\textwidth]{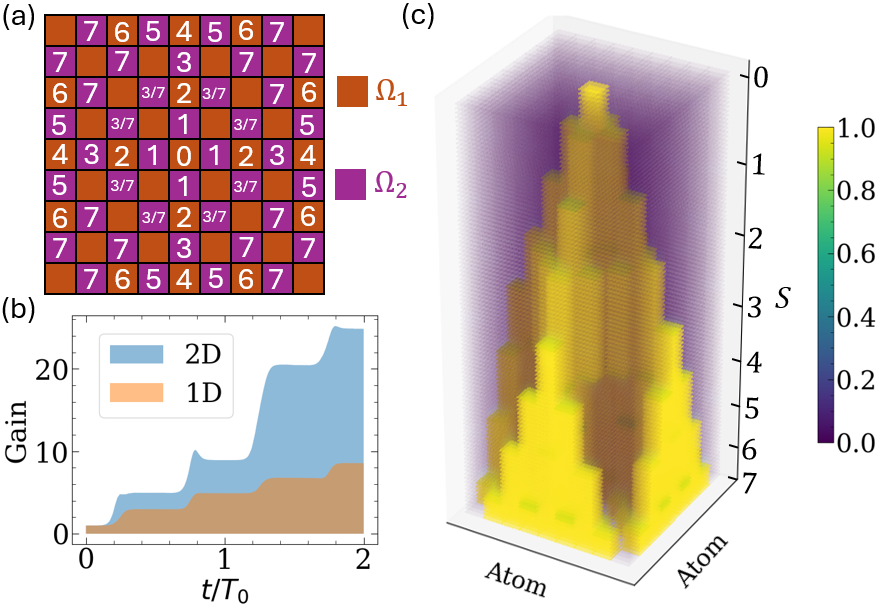}
\caption{Avalanche excitation in a 2D square lattice. (a) Atoms are located at the centers of plaquettes instead of at the vertices. Sites driven during the first (second) half-cycle are drawn in purple (brown). The site numbered ``0" is the location of the initial seed. Other numbers denote the step number during which that atom is excited. The sites labeled as ``3/7" are excited in step 3, de-excited in step 5 and re-excited in step 7. (b) Gain vs time for 1D (taken from Fig.~\ref{fig:avalanche}(b)) and 2D (MF-QMC simulations, averaged over 100 quantum trajectories). (c) Stacked images of Rydberg population evolution up to 7 steps for one trajectory that does not encounter any decay event. }
\label{fig:2D}
\end{figure}

Motivated by experimental feasibility, we next consider a 2D square lattice. A 2D geometry is advantageous because
(i) it is more practical to scale up atom numbers compared to 1D chains, (ii) the initial sensing signal has a larger receiving area, (iii) the avalanche gain per unit time is greater, and (iv) Rydberg decay is less detrimental, leading to a higher maximum gain.
In our 2D simulations, each atom interacts with its 8 nearest and next-nearest neighbors. We apply driving lasers in alternating checkerboard patterns (Fig.~\ref{fig:2D}(a)), analogous to the even-odd site driving in 1D. The parameters follow the same ramping curves (Eqs.~(\ref{eq:1})(\ref{eq:2})), but with a shifted center detuning of the ramp: $\Delta_0=V_r+V_r/(\sqrt{2})^6=22.5$~MHz, chosen as a middle-ground value to balance the different possible NNN interaction shifts. 
If the initial seed is not driven in the first step, the excitation spreads to its four nearest neighbors. If instead the seed is driven in the first step, no population transfer occurs in that step; from the second step onward, the growth is identical to the case where the seed is not driven first, just as in the 1D dynamics.

As shown in Fig.~\ref{fig:2D}(a)(b), the gain per step in 2D is at least twice as large as in 1D. Moreover, the 2D geometry is more resilient to decay. In 1D, an excitation adjacent to a hole has only one neighboring excitation. This one neighbor provides the interaction shift that moves the transition into the RAP sweeping range, enabling de-excitation and hole growth. In 2D, however, an excitation adjacent to a hole may have multiple excited neighbors, whose combined interaction shifts move the transition outside the RAP sweeping range, thereby suppressing avalanche hole growth.
Consequently, at the end of 4 steps, the gain with decay reaches 99.5\% of the corresponding no-decay gain, as compared to 95\% for 1D.  
More accurately, this high resilience is not only due to suppressed hole growth: frozen sites can become flippable through decay. This process enhances gain and further inhibits hole expansion. From a many-body physics perspective, these dynamics may offer new insights into driven, dissipative systems. We briefly discuss this in the Supplementary Information and leave a full exploration to future work.

Setting aside decay, the coherent dynamics alone exhibits richer spatial-temporal patterns in 2D. 
The cluster growth along the main axes (the vertical and horizontal line ``432101234") is compact, filling all sites along the line. Along other directions, sites get excited and de-excited multiple times because the neighboring sites are not all filled. If the array is big enough so that the excitation patterns do not change due to boundary effects, all the sites labeled ``5", ``6" and ``7" will be de-excited and re-excited in the following steps. Within the 7 steps evolved here, only the sites labeled ``3/7" undergo excitation-de-excitation cycles. The complete Rydberg population evolution pattern presented in Fig.~\ref{fig:2D}(a) is verified by evolving a single decay-free quantum trajectory (Fig.~\ref{fig:2D}(c)). If the goal is to excite every site, one may consider ditching our generic pulse sequence which naturally extends from the 1D geometry. 
Careful engineered sequences, such as step-specific ramping curves, could further enhance gain while also suppressing correlated hole growth and excitation-de-excitation cycles.

\section*{Discussion}
In conclusion, we have introduced RAP as a practical route to realize robust Rydberg antiblockade in neutral-atom arrays. By sweeping through the interaction-shifted resonance, our approach circumvents the sensitivity to position disorder and laser parameter variations that hinder antiblockade-based protocols. Numerical simulations confirm that excitation transfer remains unaffected under realistic experimental imperfections, with performance ultimately limited only by Rydberg decay. Our simulations adopt a blackbody-radiation-limited Rydberg decay at room temperature, which can be suppressed in cryogenic environments~\cite{zhang2025high,PhysRevApplied.22.024073}.
We compare our RAP scheme with a resonant Rabi protocol and find similar timescales after preliminary optimization, resulting in comparable Rydberg decay for both. The timescale of RAP is constrained by available laser power while that of Rabi is constrained by position disorder. Importantly, the Rabi scheme suffers substantial degradation from position disorder even when cooled close to the quantum limit, whereas RAP is immune to these imperfections at temperatures achievable through simple cooling techniques. 

As an example of application, we show avalanche excitation growth as a high-gain, low-dark-count amplification mechanism for sensing. In 2D geometries, the excitations grow more efficiently. 
Our quantum-trajectory analysis reveals rich correlated excitation and hole dynamics. Future work will explore emergent functionalities of this driven, dissipative, kinetically constrained system. Beyond the square lattice, other geometries, particularly frustrated ones, may offer additional opportunities.
Finally, RAP-enabled antiblockade extends well beyond avalanche excitations. For example, entangling gates and entanglement transfer protocols may be engineered by echoing two identical RAP pulses~\cite{xue2024high}.


\section*{Acknowledgments}

We thank Qi Zhou and Ian B. Spielman for insightful discussions. This work was supported by Purdue startup fund and AFOSR Grant FA9550-22-1-0327.

\bibliography{references}

\clearpage
\renewcommand{\thefigure}{S\arabic{figure}}
\renewcommand{\thetable}{S\arabic{table}}
\renewcommand{\theequation}{S\arabic{equation}}
\setcounter{figure}{0}
\setcounter{table}{0}
\setcounter{equation}{0}

%
%
%
%
%
%
%


\onecolumngrid

\section*{Supplementary Information}

\subsection*{Two-photon Rydberg excitation}
The Rydberg excitation can be realized through a one-photon excitation to a P state, or more commonly through a two-photon process via a low-lying intermediate state 5P or 6P. The one-photon excitation avoids the added decay due to the admixture of the intermediate state, but does not alleviate the demanding optical power due to the combination of weaker dipole matrix element and unfavorable wavelength. Morever, P state is more sensitive to S state to stray electric field by a factor of 6 for 70S and 70P. As a result, we continue our discussion assuming a two-photon transition coupling to an S Rydberg state. The detuning from the intermediate state must be sufficiently large such that the increased decay rate due to the admixture of the intermediate state does not derail the transport fidelity. The choice of $\Gamma/\Omega=0.2\%$ in simulations is motivated by an excitation scheme detuned from 6P by roughly 1~GHz, in combination with $\Omega/(2\pi)=3$~MHz. 

\begin{figure}[ht]
\centering
\includegraphics[width=0.5\textwidth]{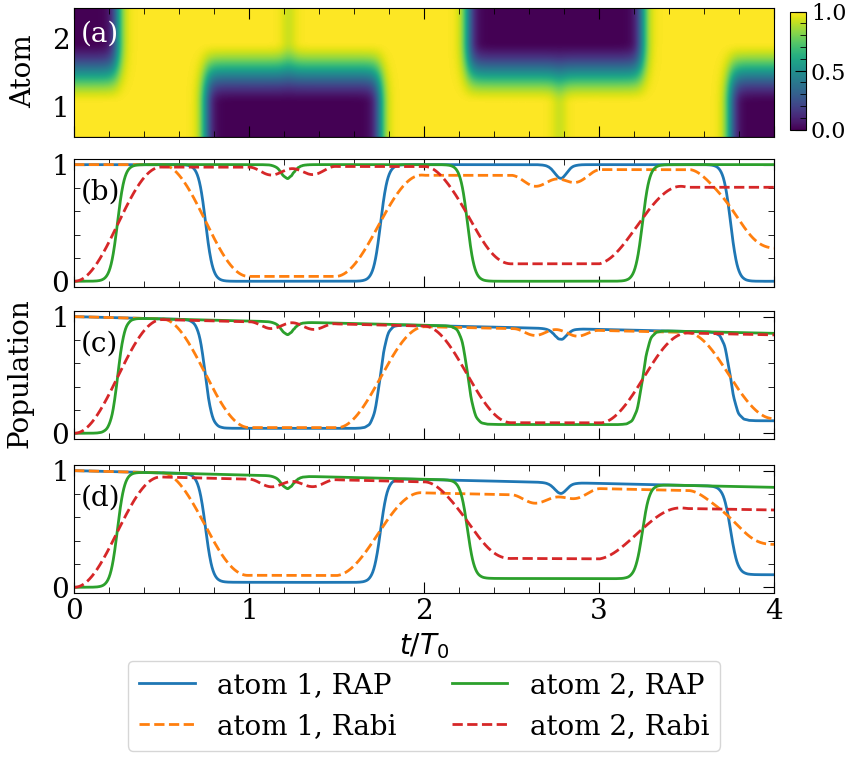}
\caption{Impacts of Rydberg-state decay and position disorder for the RAP (solid lines in (b-d)) and Rabi (dashed lines in (b-d)) schemes. (a) Rydberg population evolution without position disorder or decay. Evolution subject (b) to position disorder, (c) to decay $\Gamma=6$ MHz, and (d) to both effects. Parameters are $\Omega_0/(2\pi)=\{32,1\}$~MHz, $\sigma_r=\{54,34\}$~nm, $\overline{\delta V_{r}}/(2\pi)=\{3.06,0.10\}$~MHz, $V_r/(2\pi)=\{50,4\}$~MHz and $r=\{5.1,7.7\}~\mu$m for the RAP and Rabi schemes, respectively.}
\label{fig:disorder_decay_SI}
\end{figure}

Although the resulting fidelity from both methods are worse due to the larger decay, the overall performance of RAP is still better than Rabi. Moreover, with systematic optimization, we might be able to operate RAP less adiabatic so overall the fidelity is better.

\subsection*{1D Decay Dynamics}
Intuitively, one might expect decay to play only a minor role in avalanche dynamics. In this picture, the resulting Rydberg population distribution is simply given by $(2S+1)(1-e^{-\Gamma t})$, where $2S+1$ is the total number of excitations and each excitation has the same fidelity $1-e^{-\Gamma t}$ due to Rydberg facilitation. However, the mean-field master equation (MF-ME) approach instead produces a smooth profile in which sites farther from the seed exhibit a higher probability of being in the Rydberg state. This discrepancy arises because MF-ME fails to capture the essential facilitation physics: each new excitation should occur with probability close to its facilitated neighbor instead of unity.

Moreover, the decay of a single atom can trigger a “ground-state avalanche.” The MF-ME treatment cannot reproduce this effect, since decay in that framework only slightly reduces the average Rydberg population and therefore weakly modifies the effective interaction with neighboring sites. As a result, the MF-ME dynamics remain largely unchanged, reflecting the robustness of the adiabatic evolution to small perturbations.

\begin{figure*}[ht]
\centering
\includegraphics[width=1\textwidth]{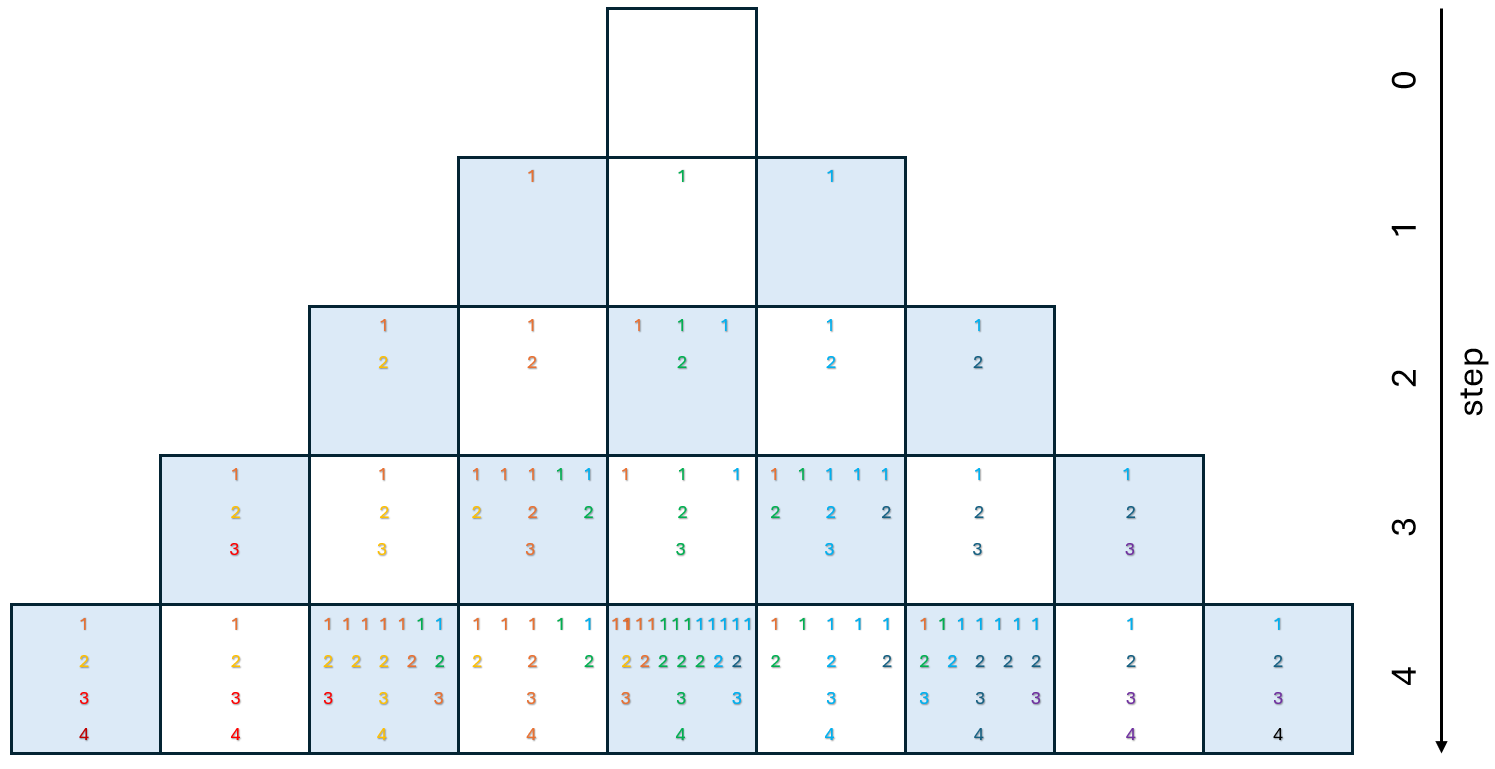}
\caption{Effect of decay in the avalanche process. Each tile represents a Rydberg excitation, and the chain grows by two sites after each adiabatic ramp (step). The blue background marks the site that is driven during the corresponding step. Colored numbers indicate the propagation of decay events: the number specifies the step at which the decay occurs (e.g., “1” corresponds to a decay in the first step), and different colors distinguish independent decay pathways. Identical characters denote sites originating from the same decay event. The Rydberg population is estimated by counting the occurrences of colored numbers; tiles with more colored numbers have a higher probability of being occupied by a “hole” rather than a Rydberg excitation.}
\label{fig:understanding of decay}
\end{figure*}

To faithfully simulate large-scale avalanche effects in the presence of decay, we instead employ a mean field quantum Monte Carlo (MF-QMC) approach rather than the master equation. The MF-QMC results agree well with the full master equation simulation, and the intuition behind the method is illustrated in Fig.~\ref{fig:understanding of decay}. Because the drive alternates between even and odd sites, the impact of a decay event depends on whether the decayed site or its neighbors are driven in the next cycle. For instance, if decay occurs on a Rydberg excitation that will be directly driven in the following cycle, its neighbors remain unaffected, and the decay appears only as a single-site “hole.” In contrast, if two neighbors of the hole are driven in the next cycle, they satisfy the condition for de-excitation (since only one excited neighbor is present), and the hole expands into these sites.

With this picture, we can estimate the site-resolved Rydberg population after several adiabatic cycles by summing over all possible decay pathways, as shown in Fig.~\ref{fig:understanding of decay}. The resulting distribution does not follow the naive understanding of “each excitation has the same fidelity $1-e^{-\Gamma t}$” or the MF-ME trend. Instead, it closely matches the MF-QMC and the full ME simulations presented in Fig.~5 of the main text. In particular, the oscillatory behavior observed in Fig.~5(b) aligns with the number of decay events shown in the fourth and fifth rows of Fig.~\ref{fig:understanding of decay}.

To obtain the gain equation [Eq.~(12)] in the main text we make the following assumptions and approximations. Assumption: at most one decay event occurs per quantum trajectory (i.e., zero or one decay during the considered time window). Approximation 1: the decay rate is small, so the decay probability is linear in time; define
\begin{equation}
    p\equiv\frac{\Gamma T_0}{2}\ll1
\end{equation}
which is the decay probability per step. Approximation 2: decays occur discretely at the end of a step and therefore the probability of a decay at a given step is $p$. With these assumptions the conditional probability that the single decay (if any) occurs at step $i$ is $(1-p)^{i-1}p$ for $1\leq i \leq S$.

For a given final step $S$, the number of ground-state atoms (holes) depends on whether a decay has occurred in any previous step. For example, if no decay occurs the gain is $G_0=2S+1$, so for $S=4$ one has $G_0=2\times4+1=9$. If the decay occurs at step $S=1$, the per-site probability is $p$ and the total number of hole avalanches growth to step $S=4$ is $7+5+7$ (the three terms denote the different site-types/pathways shown in the Fig.~\ref{fig:understanding of decay}). Similarly the probability for a decay at step $S=2$ is $(1-p)p$ with total hole-count $5+3+5+3+5$, at $S=3$ it is $(1-p)^2p$ with total hole-count $3+1+3+1+3+1+3$, and at $S=4$ it is $(1-p)^3p$ with total hole-count $9$.

Summing over all possibilities yields
\begin{equation}
    G=(2S+1)\left[1-\left(1-p\right)^{S-1}p\right]-p\sum_{i=1}^{S-1}\left[3(i+1)+i+2(2i+1)(S-i-1)\right]\left(1-p\right)^{i-1}
\end{equation}
where $p=\frac{\Gamma T_0}{2}$. Approximation 3: followed by the first approximation, we set $(1-p)\approx1$. This gives
\begin{equation}
    G=2S+1-p\sum_{i=1}^{S-1}(2S-2i+4iS-4i^2+1)=2S+1-\frac{\Gamma T_0}{2}(S^2+\frac{2}{3}S^3-\frac{2}{3}S-1)
    \label{eq:gain}
\end{equation}
The cubic term $\propto S^3$ captures the large-scale “ground-state avalanche" growth induced by a single decay. Equation~(\ref{eq:gain}) provides a good approximation to the simulations for small $S$, where the single-decay assumption is valid. A notable discrepancy appears near $S=8$: for larger $S$, the analytical expression underestimates the gain because it neglects the possibility of multiple decay events, which become increasingly likely over longer times.

\begin{figure*}[ht]
\centering
\includegraphics[width=1\textwidth]{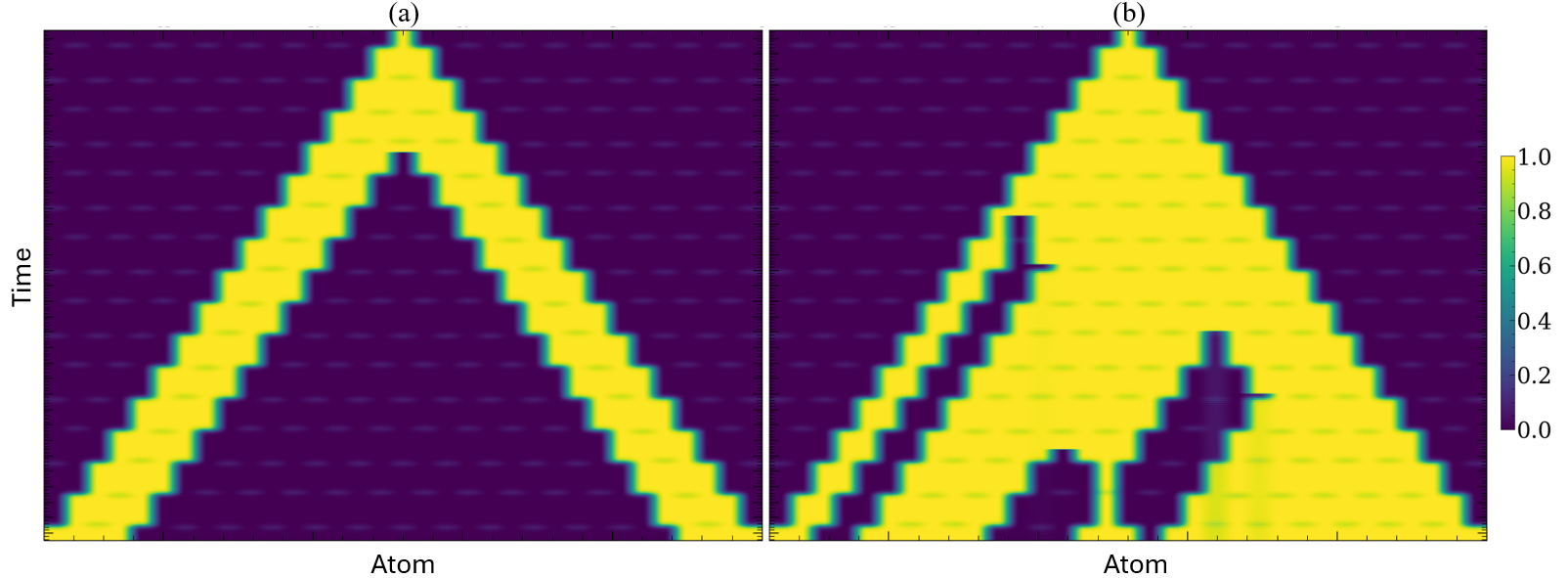}
\caption{Two single-trajectory simulations for a 33-atom system using the MF-QMC method. (a) Example of a hole-growth trajectory. (b) Example where multiple secondary decays suppress the hole avalanche.}
\label{fig:33_single_traj}
\end{figure*}

To illustrate this effect, Fig.~\ref{fig:33_single_traj} shows two single-trajectory results from the MF-QMC simulation. In the first case, with only a single decay during the avalanche, the outcome is simply the expansion of a single hole—exactly as illustrated by the green “4” in Fig.~\ref{fig:understanding of decay}. By contrast, the second trajectory demonstrates richer dynamics: the growth of the left and right holes are halted by secondary decays, while the the growth of the central hole stops when it collides with the right hole. These processes, absent from the analytical treatment, increase the overall number of excitations and explain the underestimation in Eq.~(12) compared to the simulation results in Fig.~6(a).

\subsection*{2D Decay Dynamics}

The decay dynamics in a two-dimensional array are considerably more complex than in the one-dimensional case. Even under the same assumption of a single decay per trajectory, two additional processes increase the relative weight of decay compared to non-decay events. First, the growth of a hole avalanche can become spatially confined. Second, new avalanche branches may emerge, further enriching the dynamics.

\begin{figure*}[ht]
\centering
\includegraphics[width=1\textwidth]{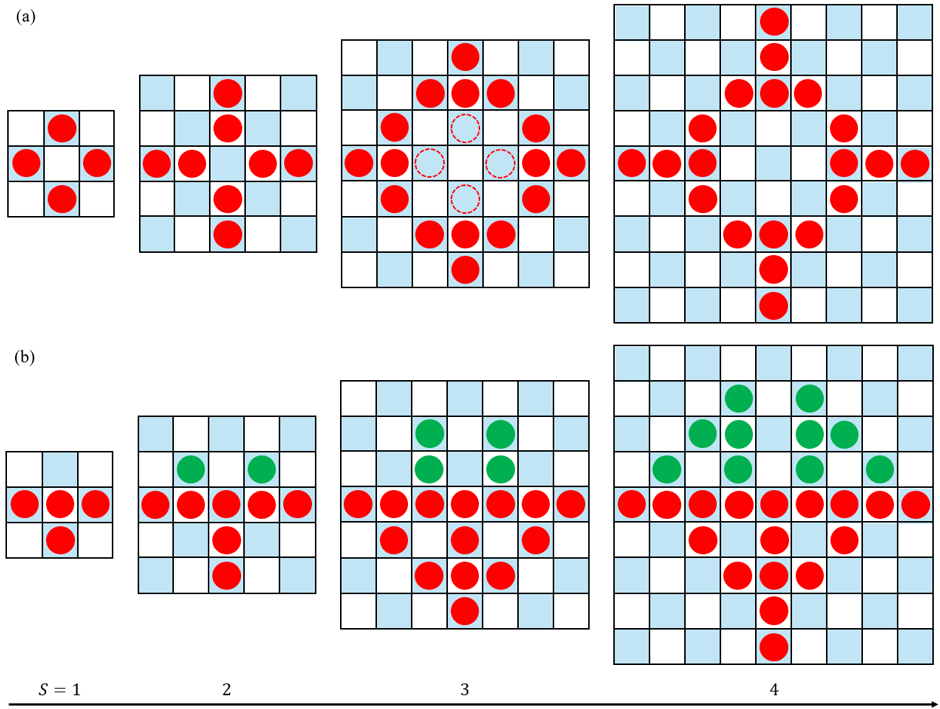}
\caption{Two single-trajectory demonstrations for a 2D array. Each tile represents a site containing one atom. A solid circle denotes a Rydberg excitation, while an empty or dashed circle denotes a ground-state atom. The blue background indicates the site driven at the corresponding step. (a) Trajectory where the central Rydberg atom decays at the end of $S=1$. (b) Trajectory where the top Rydberg atom decays at the end of $S=1$.}
\label{fig:2d_traj}
\end{figure*}

Fig.~\ref{fig:2d_traj} illustrates both scenarios. In (a), red dashed circles indicate Rydberg atoms that are adiabatically driven back to the ground state at that step. The avalanche growth of the holes halts at $S=4$ because each atom has more than one nearest neighbor, which limits further expansion. In (b), green circles denote atoms in newly formed branches. Here, the decay of a single atom generates new branches rather than triggering the expansion of a hole avalanche.

%


\end{document}